\documentclass[aps,preprint,showpacs,floatfix]{revtex4}
\usepackage{epsfig}
\usepackage{isolatin1}
\usepackage{makeidx} \makeindex
\usepackage{amsmath} \usepackage{amssymb} \usepackage{amsthm}
\usepackage{mathrsfs}

\newcommand{\ket}[1]{\,| \, {#1} \,\rangle  \,}
\newcommand{\bra}[1]{\,\langle \, {#1} \,|  \,}
\newcommand{\tgh} {{\rm tgh} \,} 
\begin{document}
\title{Decoherence Due to Discrete Noise in Josephson Qubits}
%
%
%
%
%
\begin{abstract}
We study decoherence produced by a discrete environment on a 
charge Josephson qubit by introducing a model of an environment of bistable
fluctuators. In particular we address the effect of $1/f$ noise where
memory effects play an important role.
We perform a detailed investigation of 
various computation procedures (single shot measurements, repeated 
measurements) and discuss the problem of the information needed to
characterize the effect of the environment.
Although in general information beyond the power spectrum is needed, 
in many situations this results in the knowledge of only one more 
microscopic parameter of the environment. This allows to determine which 
degrees of freedom of the environment are effective sources of decoherence 
in each different physical situation considered.
\end{abstract}
\author{E. Paladino} \affiliation{NEST-INFM $\&$ Dipartimento di Metodologie 
Fisiche e Chimiche (DMFCI), \\
        Universit\`a di Catania,
        viale A. Doria 6, 95125 Catania, Italy $\&$ INFM UdR Catania}
\author{L. Faoro} 
\affiliation{Institute for Scientific Interchange (ISI) $\&$ INFM, 
	Viale Settimo Severo 65, 10133 Torino, Italy.}
\author{G. Falci} \affiliation
{NEST-INFM $\&$ Dipartimento di Metodologie 
Fisiche e Chimiche (DMFCI),\\
        Universit\`a di Catania,
        viale A. Doria 6, 95125 Catania, Italy $\&$ INFM UdR Catania}

%
%
%
\maketitle              


A high degree of quantum coherence is crucial 
for operating quantum logic devices~\cite{kn:qcomp}.
Solid state nanodevices seem particularly
promising because of integrability and flexibility in the design
and several possible implementations have been 
proposed~\cite{kn:loss,kn:rmp,kn:teor,kn:makhlin99,kn:nature}. Few   
recent experiments succeeded in 
detecting coherent dynamics in superconducting 
devices~\cite{kn:Nakamura1,kn:vion,kn:chiorescu,kn:two-qubit}, but revealed 
limitations in the performances, due to decoherence.

In a quantum logic device the interesting degrees of freedom
are related to a given set of observables which we can
prepare (write) and measure (read), and define the {\em system}. 
Their eigenstates 
$\ket{\{q_i\}}$ form the {\em computational basis}. 
The dynamics of a state
$
\ket{\psi, t} = \sum_{q_1 \dots q_N} c_{q_1 \dots q_N}(t) 
\; \ket{{q_1, \dots, q_N}}
$
may be controlled if the Hamiltonian of the system is tunable.
Loss of coherence is due to the fact that the Hilbert space of the device is
much larger than the computational space~\cite{kn:zurek,kn:palma96}.
The additional degrees of freedom define the 
{\em environment}, which cannot be controlled and moreover little information 
is available on it. 
Decoherence is ultimately due to the entanglement between system and 
environment~\cite{kn:zurek}, but in order to pursue this point of view one 
should be able study the full dynamics, in particular system-environment 
correlations. A less fundamental point of view, which we adopt here, 
is to associate decoherence to the loss of fidelity of a quantum gate.
The environment blurs the output signal because it produces uncertainties in 
the phase relation between the amplitudes, $c_{\{q_i\}}(t)$. 

The environment degrees of freedom may be ``internal'' to the device,
which is a many-body object, or ``external'', belonging to the 
circuitry and to auxiliary devices. Internal decoherence is 
a serious problem in solid state nanodevices, 
due to the presence of many low energy 
excitations.

Investigation of the reduced qubit dynamics requires information about 
the environment. In several cases (weakly coupled environment~\cite{kn:cohen}, 
harmonic oscillator environment~\cite{kn:weiss,kn:leggett}) the 
information is contained in the power spectrum of the 
environment operators coupled to the qubit\footnote{In this spirit the 
Caldeira Leggett model has been proposed to describe quantum phenomena in 
superconducting nanocircuits~\cite{kn:dissipative}.}.
In this work we study a more general situation, an environment of 
quantum bistable fluctuators which may have memory
on the time scale of the qubit dynamics, and may display the 
effect of non-gaussian correlations.
A physical example are charged impurities in substrates and oxides.
These background charges (BC) produce for instance
the $1/f$ noise~\cite{kn:weissman} observed in metallic tunnel junctions~\cite{kn:zorin,nakamura-echo}. 

We will present here analytic results which elucidate
several aspects of the BCs environment. 
In particular we show that  decoherence 
depends on details of the dynamics of the environment beyond 
the power spectrum~\cite{kn:PRL}. 
As a consequence it may differ for different gates,
but in many cases the additional 
information required reduces to one microscopic parameter~\cite{kn:Varenna}.

\section{Model for the System and the Environment}
Superconducting qubits~\cite{kn:rmp,kn:teor,kn:makhlin99,kn:nature} are 
the only solid state implementations where coherence in a single qubit 
has been 
observed in the time domain~\cite{kn:Nakamura1,kn:vion,kn:chiorescu} 
and two-qubit systems are investigated~\cite{kn:two-qubit}. Thus  
problems on decoherence can be posed in a realistic perspective. 
In all the above implementations it has been recognized that 
$1/f$ noise is a major source of decoherence. We will focus on
charge-Josephson qubits~\cite{kn:makhlin99}, where  $1/f$ noise produced 
by BCs trapped close to the device (Fig. \ref{fig:cjqubit}).
We describe these BCs by introducing an impurity model. Many of our
results and methods can be applied to different noise sources
(eg. flux noise in flux-Josephson qubits~\cite{kn:mooji,kn:clark}) and 
other solid-state implementations (eg. gates based on the Coulomb interaction
in semiconductor-spin qubits),
but we will not discuss these topics here.

\subsection{The Superconducting Box}
\begin{figure}
\includegraphics[width=0.4\textwidth]{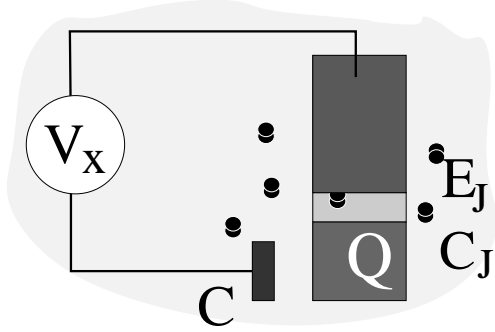}
\caption{A charge Josephson qubit in the presence of BCs located in
the substrate or in the oxide close to the junction. Relevant
scales are the charging energy $E_C=  e^2/2(C_1+C_2)$ and 
the Josephson energy $E_J$}
\label{fig:cjqubit}
\end{figure}

The charge-Josephson qubit~\cite{kn:makhlin99}  is a superconducting island
connected to a circuit via a Josephson junction and a capacitance 
$C_2$ (Fig. \ref{fig:cjqubit}). The computational states are associated
with the charge $Q$ in the island. They are mixed by the Josephson tunneling.
Level splittings can be tuned 
by the external voltage $V_x$ and 
under suitable conditions 
($E_C \gg E_J$ and low temperatures $k_B T \ll E_J$ )
only two charge states are important, which define the 
$z$ component of a pseudospin. The  
the system implements a qubit with Hamiltonian
\begin{equation}
\label{eq:qubit-hamiltonian}
{\cal H}_Q \,=\,  \, \frac{\varepsilon}{2} \, \sigma_z  -
 \frac{E_J}{2} \,\sigma_x \qquad ; \qquad 
\varepsilon(V_x) =4 E_{C} (1-C_2V_x/e)
\end{equation}
\subsection{System Plus Environment Hamiltonian}
We describe each BC as a 
localized impurity
level connected to a fermionic band~\cite{kn:PRL,bauernschmitt93,kn:mahan}. 
For a single impurity
the total 
Hamiltonian is
\begin{eqnarray}
	{\cal H}
	= {\cal H}_Q - {v \over 2}  \,
	b^{\dagger} b  \sigma_z   + {\cal H}^I \;\; ; \;\;
	{\cal H}^I =  \varepsilon_{c} b^{\dagger} b +
		\sum_{k} [T_{k}  c_{k}^{\dagger} b + \mbox{h.c.} 
]
		 +  \sum_{k} \varepsilon_{k}  c_{k}^{\dagger} c_{k} \quad\;\; 
\label{eq:hamiltonian-backcharges}
\end{eqnarray}
Here ${\cal H}^I$ describes the BC alone: 
$b$ ($b^{\dagger}$) destroys (creates) an electron 
in the localized  level $\varepsilon_{c}$ and
the electron may tunnel with amplitude $T_{k}$ to a band,
described by the operators $c_{k}$, $c^{\dagger}_{k}$ and the energies 
$\varepsilon_{k}$. 
An important scale is the switching rate 
$\displaystyle{\gamma=2 \pi {\cal N}(\epsilon_{c})|T|^2}$
(${\cal N}$ is the density of states of the band, and 
$|T_{k}|^2 \approx |T|^2$), which characterizes the relaxation 
regime of the BC. The BC determines a bistable extra bias $v$ for the qubit,  
via the coupling term.
For a set of 
BCs we generalize Eq. (\ref{eq:hamiltonian-backcharges}) as follows
\begin{equation}
	{\cal H}
	= {\cal H}_Q - {1 \over 2}  \,\hat{E} \,
	\sigma_z   + \sum_i {\cal H}^I_i
\label{eq:hamiltonian-backcharges2}
\end{equation}
where extra bias operator is 
$\hat{E}(t)=\sum_i v_i b^\dagger_i b_i$. For simplicity the 
assumption that each localized level is connected to a distinct band has been 
made.
\subsection{Model for $1/f$ Noise}
\label{sec:model-oneoverf} 
The environment in Eq. (\ref{eq:hamiltonian-backcharges2}) above
is specified by the distribution of the switching rates $\gamma_i$ which
we choose in order to reproduce the $1/f$ noise.
The standard way~\cite{kn:weissman} is to assume a distribution 
$P(\gamma) \propto 1/ \gamma$ for $\gamma \in [\gamma_m,\gamma_M]$.
Indeed if we look at the relaxation regime of the BC, the 
total extra polarization is a classical stochastic process 
$E(t)$ with power spectrum 
\begin{equation}
\label{eq:powerspec}
S(\omega)\;=\; \sum_i S_i(\omega) \qquad ; \qquad 
S_i(\omega)\;=\; {1 \over 2} \, v_i^2 \,  
(1- \overline{\delta p}^2) \; {\gamma_i \over (\gamma_i^2 + \omega^2)}
\quad,
\end{equation}
$\overline{\delta p}$ being the thermal average of the 
difference in the populations of the 
two states of the BC, the distribution $P(\gamma)$ leads to to $1/f$ noise, 
$S(\omega)= 
\{\pi (1- \overline{p}^2)\,  n_d \,\overline{v^2} /(4 \,\, \ln \, 10)\} \,\,
\omega^{-1}
$
for frequencies  
$\omega \in [\gamma_m,\gamma_M]$  ($n_d$ is the number of
fluctuators per noise decade). With the above choice of 
parameters, the Hamiltonian Eq. (\ref{eq:hamiltonian-backcharges2}) 
has been used to study decoherence due to $1/f$ noise~\cite{kn:PRL}. 
 
\section{Reduced Dynamics} 
\label{sec:reduced-dynamics}
Our aim is to investigate the effect of the BC 
environment on the dynamics of the qubit.
The standard road-map is to calculate the reduced density 
matrix~\cite{kn:weiss} of the qubit 
$\rho^Q(t) = \mathrm{Tr}_E \{W(t)\}$, $W(t)$ being the full 
density matrix. In the standard weak coupling approach a master equation 
for $\rho^Q(t)$ is written~\cite{kn:cohen}, the environment 
entering its dynamics only via the power spectrum, Eq. (\ref{eq:powerspec}). 
A master equation for $\rho^Q(t)$ 
can also be obtained by modeling the environment by a set of 
harmonic oscillators. By using diagrammatic 
techniques~\cite{kn:weiss,kn:grifoni,kn:shnirman} results of the standard 
weak coupling approach are obtained
at lowest order in the couplings $v$, but it has been pointed out that 
higher orders are important for a $1/f$ oscillator 
environment~\cite{kn:PRL,kn:shnirman,kn:cottet}. 

The failure of the standard weak coupling approach is due to the fact that 
the $1/f$ environment includes fluctuators which are very slow on the 
time scale of the reduced dynamics, the approach being reliable 
only for  BCs with $v_i \ll \gamma_i $~\cite{kn:cohen,kn:PRL}. 
Rather than study 
higher orders in the perturbation series~\cite{kn:shnirman}, 
we prefer to use a different strategy. We study the models 
Eqs.(\ref{eq:hamiltonian-backcharges},\ref{eq:hamiltonian-backcharges2}) by 
enlarging the system and considering only the bands as the environment. 
The modified road-map consists in calculating in some approximation 
the reduced density matrix (RDM) 
$\rho(t) = \mathrm{Tr}_{bands} \{W(t)\}$ and then extract exactly 
$\rho^Q(t)$. This allows to obtain results to 
all orders in the coupling $v$ and to investigate details of various quantum 
gates. The disadvantage is that we have a larger system to deal with. We 
have investigated this problem with different techniques~\cite{kn:PRL}, 
here we present a master equation approach, 
which covers many of the results we obtained before.

\subsection{Master Equation for a Single BC}
We split Eq. (\ref{eq:hamiltonian-backcharges}) in system, 
 ${\cal H}_0={\cal H}_Q - {v \over 2}  \,
	b^{\dagger} b  \sigma_z + \varepsilon_{c} b^{\dagger} b$, and 
environment, 
${\cal H}_E = \sum_{k} \varepsilon_{k}  c_{k}^{\dagger} c_{k}$, coupled 
by
${\cal V} = \sum_{k} [T_{k}  c_{k}^{\dagger} b + \mbox{h.c.} ]$. 
The eigenstates 
(and eigenvalues) of ${\cal H}_0$ are 
product states of the form $\ket{qubit}\ket{BC}$, namely
$\ket{a} = \ket{\theta +}\ket{0}$ 
($ -  \Omega/2 $), 
$\ket{b} = \ket{\theta -}\ket{0}$ 
($\Omega/2 $), 
$\ket{c} = 
\ket{\theta^\prime +}\ket{1} 
$
($-  \Omega^\prime/2  + \varepsilon_c$),
$\ket{d} = 
\ket{\theta^\prime -}\ket{1}$
($\Omega^\prime/2 + \varepsilon_c$).
Here
$\ket{\theta \pm}$ are the two eigenstates of $\sigma_{\mathbf{\hat{n}}}$,
the direction being specified by the polar angles 
$\theta$ and $\phi=0$.
Each set of qubit states corresponds to a value of 
$b^\dagger b = 0,1$: the two level splittings are 
$\Omega = \sqrt{\varepsilon^2 + E_J^2}$ and 
$\Omega^\prime = \sqrt{(\varepsilon+v)^2 + E_J^2}$, and finally
$\cos \theta = {\varepsilon / \Omega}$, $\sin \theta = {E_J / \Omega}$, 
$\cos \theta^\prime = {(\varepsilon+v) / \Omega^\prime}$, 
$\sin \theta^\prime = {E_J / \Omega^\prime}$.

In the basis of the eigenstates of ${\cal H}_0$ the master equation for the 
RDM in the Schr\"odinger representation reads
\begin{eqnarray}
\label{eq:master-equation-sch}
{d \rho_{ij}(t) \over dt} &=& - i \, \omega_{ij} \, \rho_{ij}(t) \,+\,
\sum_{mn} \;
\mathrm{R}_{ij \, mn} \; 
\;\rho_{mn}(t) 
\end{eqnarray}
where $\omega_{ij}$ is the difference of the eigenenergies and $\mathrm{R}_{ij \, mn}$
are the elements of the Redfield tensor~\cite{kn:cohen}
\begin{eqnarray}
\mathrm{R}_{ij \, mn}  &=& \int_0^\infty 
	\hskip-5pt d \tau \;  \Bigl\{ 
		C_{njmi}^> (\tau) \,\mathrm{e}^{i \omega_{mi}\tau} + 
		C_{njmi}^< (\tau) \,\mathrm{e}^{i \omega_{jn}\tau} 
\nonumber\\ && \hskip20pt - 	
		\delta_{nj} \sum_k  C_{ikmk}^>(\tau) \,
			\mathrm{e}^{i \omega_{mk} \tau} 	
		\,-\, \delta_{im} \sum_k  C_{nkjk}^<(\tau) \,
			\mathrm{e}^{i \omega_{kn} \tau} 	\Bigr\}
\qquad
\end{eqnarray}
The correlation functions are given by
$$
C^{{\raisebox{-3pt}{\tiny $>$} \atop \raisebox{2pt}{\tiny $<$}} }_{ijkl}(t) = \bigl[
\bra{i} b \ket{j} \, \bra{l} b^\dagger \ket{k}
+ \bra{i} b^\dagger \ket{j} \, \bra{l} b^\dagger \ket{k} \bigr] \; i G^{{\raisebox{-3pt}{\tiny $>$} \atop \raisebox{2pt}{\tiny $<$}} }(t)
$$
where
$
i G^>(\omega) = {\gamma /( 1 + \mathrm{e}^{-\beta \omega})}$, 
$G^<(\omega) = G^>(-\omega)$.
Many of the coefficients $\mathrm{R}_{ijmn}$ vanish. In particular the system 
of equations (\ref{eq:master-equation-sch})
splits in two blocks. The first contains the 
populations and the coherences $\rho_{ab}$ and 
$\rho_{cd}$, together with their conjugate, i.e. the elements 
diagonal in the BC. 
The second block contains all the other coherences, which
vanish identically if the initial RDM $\rho(0)$ is diagonal in the BC, 
the physically relevant case for our purposes.

In the standard secular approximation~\cite{kn:cohen} in the r.h.s. of
Eq. (\ref{eq:master-equation-sch})
are retained only terms with coefficients 
$\mathrm{R}_{ij\, ij}$, but this is not enough in 
our problem. To discuss possible approximations we first point out that 
we are interested in the case $v \ll \Omega,\Omega^\prime$. 
We have two important scales,
namely $\Omega^\prime - \Omega$ and $\Omega \sim \Omega^\prime$, the latter being much larger than the former. The secular approximation is valid if 
$\gamma \ll \Omega^\prime - \Omega$,
i.e. for an almost static BC. If 
$\gamma \sim \Omega^\prime - \Omega \ll \Omega \sim \Omega^\prime$
we can no longer neglect the coefficients mixing $\rho_{ab}$ and $\rho_{cd}$ 
(and those mixing the conjugates). We call this regime adiabatic. 
Finally in the fast BC regime, $\gamma \gtrsim \Omega, \Omega^\prime$, 
all the $\mathrm{R}_{ij\, mn}$ must be present. Despite of this,  
the reduced dynamics of the qubit is just the result of
lowest order approach \cite{kn:weiss,kn:grifoni,kn:shnirman}, 
so we will not discuss this regime anymore. 
\subsection{Results for Adiabatic BCs}
In the adiabatic regime coherences are calculated using the following elements of the Redfield tensor
\begin{eqnarray}
\mathrm{R}_{ab \, ab} = - 
{\gamma \over 2}  \bigl[ 1 - c^2  \delta - s^2   \delta^\prime 
+ i
\bigl( c^2 w + 
s^2 w^\prime
\bigr) \bigr] &\; ; \;&
\mathrm{R}_{ab \, cd}  = 
 {c^2 \gamma \over 2}
\bigl( 1  + \delta  - i w \bigr)
\nonumber\\
\mathrm{R}_{cd \, cd} = - 
{\gamma \over 2} 
 \bigl( 1 + c^2  \delta + s^2   \delta^\prime 
+ i 
\bigl( c^2 w - 
s^2 w^\prime
\bigr) \bigr]
&\; ; \;&
\mathrm{R}_{cd \, ab}  =
{c^2 \gamma \over 2}
\bigl( 1  - \delta  - i w \bigr) \quad
\nonumber
\end{eqnarray}
where $c=\cos [(\theta-\theta^\prime)/2]$, 
$s=\sin [(\theta-\theta^\prime)/2]$, 
$\delta = t_{ca} + t_{db}$, $\delta^\prime = t_{da} +  t_{cb}$, 
$w = w_{ca} - w_{db}$,
$w^\prime = w_{da} - w_{cb}$ and
\begin{eqnarray}
t_{ij} = 
\frac{1}{2} \, \tgh \Bigl({\beta \omega_{ij} \over 2} \Bigr) \quad; \quad 
w_{ij} = - {1 \over \pi}
\; \Re \Bigl\{ \psi \Bigl({\pi + i \beta \omega_{ij} \over 2 \pi}
\Bigr) \Bigr\} \, .
\nonumber
\end{eqnarray} Here $\psi(z)$ is the digamma function. 
The coherences $\rho_{ab}(t)$ and $\rho_{cd}(t)$ can be found in closed form
by simply diagonalizing a $2 \times 2$ matrix. They allow to study the
qubit coherence via 
$\langle \sigma_y (t) \rangle \;=\; - 2 \Im [\rho_{ab}(t)+\rho_{cd}(t)]$.
The qubit  coherence decays as $\exp\{-\Gamma(t)\}$, where 
\begin{eqnarray}
\label{eq:decay-coherences}
\Gamma(t) =
- \; \ln \left| {\rho_{ab}(t) + \rho_{cd}(t) \over 
\rho_{ab}(0) + \rho_{cd}(0)} \right| \, .
\end{eqnarray}

Here we present the analytic solution in a regime where the dynamics of the 
charge is not modified by the presence of the qubit, a case which 
is of interest for $1/f$ noise. We find 
\begin{eqnarray}
\label{eq:adiabatico-prima-analisi-1}
\rho_{ab}(t) + \rho_{cd}(t) =  \mathrm{e}^{i(\Omega+\gamma g/2)t} \;
{1 \over 2 \alpha} \; \bigl\{ A(\alpha) 
\, \mathrm{e}^{- {\gamma \over 2} \, (1-\alpha) t} \,-\,
 A(-\alpha) \, \mathrm{e}^{- {\gamma \over 2} \, (1+\alpha) t}
\bigr\}
\quad 
\end{eqnarray}
where
$\alpha =  \sqrt{1-g^2-2i g \overline{\delta p}- (1-c^4) (1- \overline{\delta p}^2)}$ and
 $g=(\Omega^\prime-\Omega)/\gamma$ enter the decay rates whereas the prefactors are 
$$
A(\alpha) = (\alpha + c^2 - i g^\prime )\, 
\rho_{ab}(0) + (\alpha + c^2 + i g^\prime)\, 
\rho_{cd}(0)
\;\; ; \;\; g^\prime = g + i  \, \overline{\delta p} \, (1 - c^2) \, .
$$

\subsection{Two Regimes for the BC}
A rough analysis of 
Eq. (\ref{eq:adiabatico-prima-analisi-1}) allows to draw a physical picture.
If $g \gg 1$, $\alpha$ is substantially imaginary, reflecting the fact that
a very slow BC mainly provides a static energy shift.  
Instead for $g \ll 1$ the decay rates acquire a  
real part thus a fast BC may determine an 
exponential reduction of the output signal. From a more quantitative 
analysis~\cite{kn:PRL} it emerges that 
BCs with $g \ll 1$, which we call {\em weakly coupled},  
behave as a suitable set of harmonic oscillators with 
power spectrum given by Eq. (\ref{eq:powerspec}). They mainly produce
the homogeneous broadening of the signal. Instead BCs with
$g \gg 1$, which we call {\em strongly coupled}, give rise to memory effects 
and deviations of their statistical properties from those
of an oscillator environment (cumulants higher than the second) are 
relevant~\cite{kn:weissman}. They mainly produce the inhomogeneous 
broadening of the signal.

Finally we compare this result with other approaches. Notice first that 
Eq. (\ref{eq:adiabatico-prima-analisi-1}) is 
valid if the power spectrum $S_i(\omega)$ 
(see Eq. (\ref{eq:powerspec})) is negligible at frequencies 
$\omega \sim \Omega$. In this regime
both the decay $\Gamma(t)$
Eq. (\ref{eq:decay-coherences}) and the energy shift 
reproduce the results of the 
standard  weak coupling approach~\cite{kn:cohen} if $\gamma t \gg 1$. 
On the other hand, as in the work of 
Refs.~\cite{kn:weiss,kn:grifoni,kn:shnirman}, 
we may simulate the effect of the BC by a suitable set of quantum 
harmonic oscillators. The resulting decoherence rate  for this 
oscillator environment is given by
\begin{equation}
\label{eq:gamma-osc-1}
\Gamma_{osc}(t) = {1 \over 2} \; g^2 \; \left[ {\partial^2 \Gamma(t) \over \partial g^2} \right]_{g=0}
\end{equation}
where $\Gamma(t)$ is given by Eq. (\ref{eq:adiabatico-prima-analisi-1}) 
with thermal initial conditions for the BC.

\section{Pure Dephasing}
For $E_J = 0$ the environment only produces random fluctuations of the level splitting. 
This  case, usually referred to as ``pure dephasing'', is special 
in that the Hamiltonian~(\ref{eq:hamiltonian-backcharges2}) commutes with $\sigma_z$. 
The charge in the island is then conserved and no relaxation occurs, but if
we prepare the qubit in a superposition the system will 
dephase.

If the initial density matrix is factorized, $W(0)= w_E(0) \otimes \rho^Q(0)$, 
it is possible to write an exact expression for the coherences only in terms
of the environment~\cite{kn:palma96,kn:PRL,kn:Varenna},
\begin{equation}
\label{eq:gamma-pure-dephasin-general}
\rho^Q_{01}(t) = \rho^Q_{01}(0) \, \mathrm{e}^{-\Gamma(t) - i \delta E(t)} \; ; \;
\Gamma(t) = - \ln \left|
\mathrm{Tr}_E \bigl\{ w_E(0) \,
\mathrm{e}^{i{\cal H}_{-1}t} \mathrm{e}^{-i{\cal H}_{1}t}
\bigr\} \right| \;
\end{equation}
where ${\cal H}_\eta = \sum_i {\cal H}^I_{i} - (\eta/2) \,\hat{E}$. This general
expression may be further simplified because individual charges contribute 
independently to $\Gamma(t)$, $\Gamma(t)=\sum_i \Gamma_i(t)$, where 
$\Gamma_i(t)$ is the dephasing due to a single charge resulting
from the Hamiltonian Eq. (\ref{eq:hamiltonian-backcharges}).  
We notice that in this case $\theta = \theta^\prime$, which implies 
that  $\rho_{10}^Q(t)=\rho_{ab}(t)+\rho_{cd}(t)$ and moreover the adiabatic
approximation is exact, since the Redfield coefficients relating the 
above two coherences with all the other entries of the RDM vanish identically. 
Then $\Gamma_i(t)$ is given by Eq. (\ref{eq:decay-coherences}) and in 
the limit described by Eq. (\ref{eq:adiabatico-prima-analisi-1}) 
we find the analytic form
\begin{equation}
\label{eq:gamma-pure-dephasing-singlecharge}
\Gamma_i(t) = - \ln \left|
A_i\, \mathrm{e}^{- {\gamma_i \over 2} \, (1-\alpha_i) t} \,+\,
(1- A_i) \, \mathrm{e}^{- {\gamma_i \over 2} \, (1+\alpha_i) t}
\right| \quad
\end{equation}
where 
$
A_i = [1 + (1 - i g_i)/\alpha]\,p_{0i} + [1 + (1 + i g_i)/\alpha]\,p_{1i}
$, 
$g_i = v_i/\gamma_i$, $\alpha_i = \sqrt{1-g_i^2-2i g_i \overline{\delta p}_i}$, and
$p_{0i}$ ($p_{1i}$) is the probability that the $i$-th BC is initially empty (singly occupied). 
This result has been obtained using several different techniques in 
\cite{kn:PRL,kn:Varenna}.

Finally by combining the equations of this section with 
Eq. (\ref{eq:gamma-osc-1}) we recover the exact
result of a set of harmonic oscillators~\cite{kn:palma96}
\begin{equation}
\label{eq:gamma-pure-dephasing-oscillators}
\Gamma_{osc}(t)
\;=\; \int_{0}^{\infty} \hskip-2pt {d\omega \over \pi} \;
S(\omega) \;
{1 - \cos \omega t \over \omega^2} \,. 
\end{equation}

\subsection{Single BC}
\label{sec:single-BC}
We study deviations of
Eq. (\ref{eq:gamma-pure-dephasing-singlecharge}) from
the result for an oscillator environment,
Eq. (\ref{eq:gamma-pure-dephasing-oscillators}).
We consider different initial 
conditions for the BCs. 
Substantial deviations are clearly observed, except in the case of a weakly coupled BC 
($g=0.1$). In particular BCs with $g >1$ induce slower dephasing compared 
to an oscillator environment with the same $S(\omega)$, a sort of saturation 
effect. Recurrences 
at times comparable with $1/v$ are visible in $\Gamma(t)$. 
\begin{figure}[h!]
\includegraphics[width=0.50\textwidth]{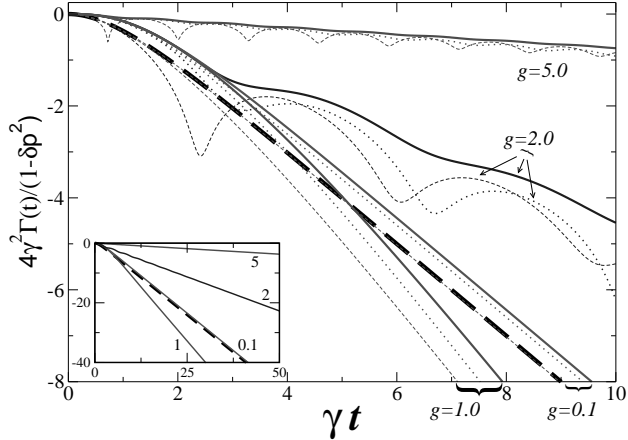}
\caption{Reduced $\Gamma(t)$ due to a BC prepared in a stable state
($\delta p_0 = 1$ solid lines and $\delta p_0=-1$ dotted lines) 
and in a thermal mixture ($\delta p_0 = \overline{\delta p}$ dashed lines)
for the indicated 
values of $g=v/\gamma$. Inset: longer time behavior 
for stable state preparation. 
The curves are normalized in such a way that the oscillator approximation for all of them coincides (thick dashed line)}
\label{fig5}
\end{figure}
In addition strongly coupled charges show memory effects. 
We now study the effect of the initial conditions of the BC, expressed via
$\delta p_0=p_0-p_1$. 
In a single shot process $\delta p_0 = \pm 1$ and $\Gamma(t)$ describes dephasing {\em during} 
time evolution (homogeneous broadening). Memory effects are apparent and 
in particular for $\gamma t \ll 1$ the short time behavior is 
$\Gamma(t) \approx  v^2 t^2 (1-{\delta p}_0^2)/8
+ \gamma v^2 t^3 \, (1+2 \delta p_0 \overline{\delta p}-3 \delta p_0^2 )/24 \propto t^3$.
A two level system is stiffer than a set of oscillators and indeed
$\Gamma_{osc}(t) \approx v^2  (1-\overline{\delta p}^2)/8 \, t^2$.
On the other hand if we choose $\delta p_0 = \overline{\delta p}$ in 
Eq. (\ref{eq:gamma-pure-dephasing-singlecharge}), for very short times 
$\Gamma(t) \approx \Gamma_{osc}(t)$. This case
corresponds physically to repeated measurements where the 
preparation of the BC is not controlled and determines slightly different 
characteristic frequency for the qubit. This sort of inhomogeneous broadening
adds to decoherence during the individual time evolutions and determines
a faster decay of $\Gamma(t)$. 

To summarize a weakly coupled BC, $g \ll 1$, behaves as a source of 
gaussian noise, decoherence depending only on the power spectrum
of the fluctuator, whereas decoherence due a strongly coupled BC, 
$g \gg 1$, displays saturation effects and dependence on the
initial conditions of the BC.

\subsection{$1/f$ Noise in Single Shot Measurements}
\label{sec:single-shot}
The set of BCs producing $1/f$ noise contains both weakly
and strongly coupled fluctuators, so no typical time scale is
present. A  very large number of slow fluctuators is present, and 
it is not a priori clear how saturation manifests.
\begin{figure}[h!]
\includegraphics[width=0.52\textwidth]{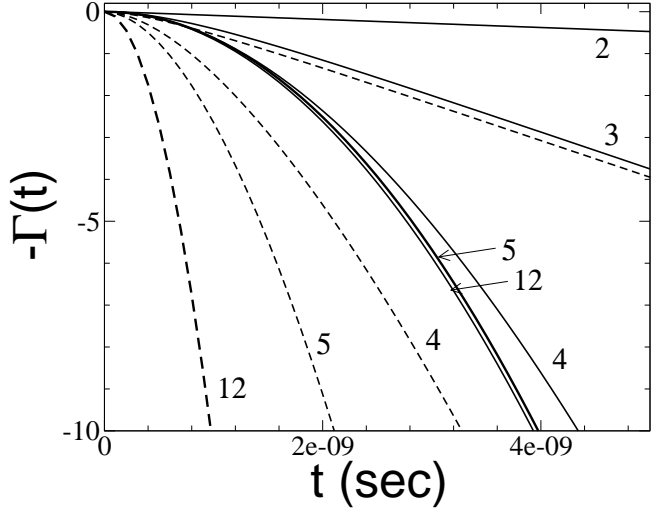}
\caption{$\Gamma(t)$ due to $1/f$ noise in the range $\omega \in [\gamma_m,10^{12}\,Hz]$, 
for decreasing 
$\gamma_m$ (solid lines, the label is the number of decades included).
Slower BCs saturate whereas this not happens for 
an oscillator environment (dashed lines).
Parameters ($\overline{v} = 9.2 \times 10^7 Hz$, $n_d = 1000$)
give the experimental noise levels
and reproduce 
the observed decay of the echo signal~\protect\cite{nakamura-echo}.
Couplings $v_i$ are distributed with  
$ \Delta v / \overline{|v|} = 0.2$}
\label{fig5b}
\end{figure}
In Fig. \ref{fig5b} we 
show the results for a realistic sample. We choose initial conditions 
$\delta p_{0j} = \pm 1$ randomly distributed on the set of $N$ BCs 
with $(1/N) \sum_j \delta p_{0j} \approx \overline{\delta p}$. We checked that 
different microscopic realizations of this conditions give roughly the
same total initial extra bias $E(0)$ and the same dephasing. We 
calculate dephasing during the time evolution, 
i.e. the average signal of several single-shot experiments, where 
$E(0)$ is re-calibrated before each experiment. This ideal protocol 
minimizes the effects of the environment. 

We now perform a spectral analysis of the effects of the environment
by adding slower fluctuators decade after decade, in a way such that
$1/f$ noise with the same amplitude $\cal A$ is present for $\omega \in 
[\gamma_m,10^{12}\,Hz]$.
We see in this example that dephasing is given by BCs 
with $\gamma_j > 10^7 Hz \approx \overline{|v|} /10$.
The overall effect of the strongly coupled BCs 
($\gamma_j < \overline{|v|} /10$) is minimal, despite of their large 
number, thus low-frequency noise saturates. Instead
$\Gamma_{osc}(t)$,  Eq. (\ref{eq:gamma-pure-dephasing-oscillators})
does not saturate at low frequencies.
We notice finally that
for this protocol $\Gamma(t)$ is roughly given by $\Gamma_{osc}(t)$
provided low frequencies are cut off at $\omega \sim \overline{|v|}$.
Thus dephasing depends essentially on 
a single additional parameter besides the power spectrum $S(\omega)$, namely 
the average coupling  $\overline{|v|}$ or equivalently the number 
of charges producing a decade of noise, $n_d$.

Our results are not very 
sensitive to the value of $n_d$ we choose. Indeed for constant 
amplitude $\cal A$ we must keep constant 
$\sum_i v_i^2 \approx n_d \overline{v^2}$, so the
effective low-frequency cutoff $\sim \overline{|v|}$ varies 
as $n_d^{-1/2}$. For $n_d \to \infty$ the low-frequency cutoff goes to
zero, as in the result Eq. (\ref{eq:gamma-pure-dephasing-oscillators}) 
for the oscillator environment.

\subsection{Repeated Measurements and Inhomogeneous Broadening}
\label{sec:inhomogeneuos}
Single-shot measurements are a goal for experimental research, but 
presently available protocols involve repeated measurements.
For instance in \cite{kn:Nakamura1,nakamura-echo} the time 
evolution  procedure
is repeated $\sim 10^{5}$ times and 
the total current due to the possible presence of the extra Cooper pair 
in all the repetitions is measured. The signal is then the sum over different possible
time evolutions of the BCs with initial conditions which are also randomly 
fluctuating. This additional blurring of the output signal results in
a faster decay for $\Gamma(t)$. As explained in \cite{kn:Varenna},
\vspace{3.5mm}
\begin{figure}[bht]
\includegraphics[width=0.50\textwidth]{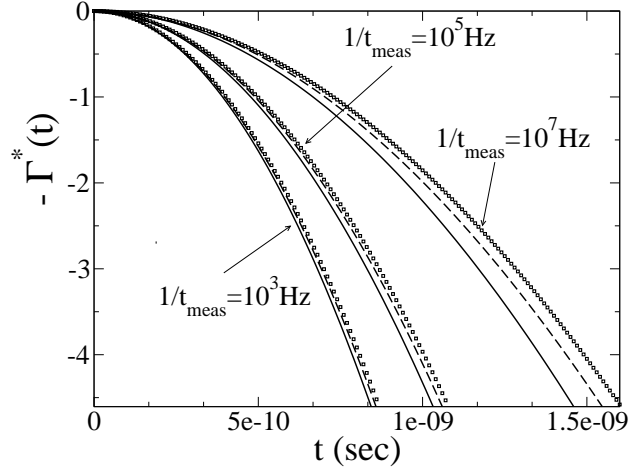}
\caption{
Different averages over $\delta p_{0j}$
for $1/f$ spectrum reproduce the effect of repeated measurements.
They are obtained by neglecting (dotted lines) or account for 
(solid lines) the strongly correlated dynamics of $1/f$ noise.
The noise level of \cite{nakamura-echo} is used, by setting
$\overline{|v|} = 9.2 \times 10^6 Hz$, 
$n_d=10^5$, $\gamma_m = 1 Hz$, $\gamma_M = 10^9 Hz$. 
Dashed lines are the oscillator approximation with 
a lower cutoff at $\omega=\min \{\overline{|v|} , 1/t_m\}$  
}
\label{gaussian}
\end{figure}
this is accounted for by letting in $\Gamma(t)$ 
(i.e. in Eqs.(\ref{eq:gamma-pure-dephasing-singlecharge})) 
$\delta p_{0i} = \overline{\delta p}_i(t_m)$,  
the average of the values of  $\delta p_j$ sampled
at regular times $t_\alpha$ for the overall measurement time $t_m$. 
As a rough estimate 
we let $\overline{\delta p}_i(t_m)= \pm 1$ 
if $\gamma_i t_m < 1$ and 
$\overline{\delta p}_i(t_m)= \overline{\delta p}_i$ 
for $\gamma_i t_m > 1$, and 
consider only the case of long overall measurement time 
$\overline{|v|} t_m \gg 1$. In this case BCs with $\gamma < 1/t_m$ are
saturated and are not effective,  whereas for 
the other BCs, being averaged, we may take 
$\Gamma^{(i)}(t) \approx \Gamma^{(i)}_{osc}(t)$ for small enough times. 
The result would be
$\Gamma(t) \approx \int_{1/t_m}^{\infty} \hskip-2pt d\omega \;
S(\omega)(1 - \cos \omega t)/(\pi \omega^2)$ and
would proof the recipe proposed by Cottet et al.~\cite{kn:cottet}. 

In Fig.  \ref{gaussian}, (dotted lines) we show that indeed 
dephasing calculated as outlined
above, is roughly given at short times by the oscillator environment 
approximation with a lower cutoff taken at 
$\omega \approx  \min \{\overline{|v|}, 1/t_m\}$. 
We also show results with a different averaging procedure 
$\delta p_j^0(t_m)= 1/t_m \int_0^{t_m} dt 
\overline{\delta p(t)}$
which takes 
into account the strongly correlated dynamics of $1/f$ noise 
(solid lines in Fig.  \ref{gaussian},). 
These correlations do not affect the results except possibly for 
$t_m \approx \overline{|v|}$.

\subsection{Charge Echo}
Echo-type techniques have been recently suggested~\cite{kn:Nakamura1,kn:cottet} and 
experimentally tested~\cite{nakamura-echo} as a tool to reduce 
inhomogeneous broadening due to the low-frequency fluctuators of the 
$1/f$ spectrum. In the experiment of ~\cite{nakamura-echo} the echo 
protocol consists of a $\pi/2$ preparation pulse, 
a $\pi$  swap pulse and a $\pi/2$ measurement pulse. Each pulse is separated
by the delay time $t$.
We calculated the decay of the echo signal using a semi-classical 
approach~\cite{kn:mqc2,kn:Varenna}. This result allowed to estimate the
parameters we have used in this work by comparing with ~\cite{nakamura-echo}.

The decay  depends very weakly on 
initial conditions of the BCs and is well reproduced by the oscillator environment
approximation. This means that in the experiment~\cite{nakamura-echo} the echo procedure 
actually cancels the effect of 
strongly coupled charges, this conclusion being valid as long as 
the delay time is short, $t \overline{|v|} \ll 1$.
We remark that in the regime of parameters we
consider, for given noise amplitude the echo signal is strongly dependent 
on the high frequency cutoff and a detailed analysis may give information
on the actual existence of BCs switching at rates comparable with 
$\Omega \sim 10$ GHz.

\section{Decoherence for a Generic Working Point}
The most effective strategy for defeating $1/f$ noise has been implemented
in the experiment by Vion et al.~\cite{kn:vion}. It is useful to explain it in a non pictorial
way by considering decoherence as given by the weak coupling 
approach~\cite{kn:cohen,kn:weiss,kn:grifoni}
\begin{equation}
\label{eq:decoherence-golden}
\Gamma_0(t) = \Bigl[{1 \over 2} \, S(0) \, \cos^2\theta + 
{1 \over 4} \, S(\Omega) \, \sin^2\theta \Bigr] \; t \, .
\end{equation}
Even if this formula does not hold for $1/f$ noise, it indicates that the dangerous
adiabatic term containing $S(0)$ may be eliminated (in lowest order) if one 
operates at $\theta = \pi/2$. Small deviations from this point may
determine a dramatic increase of decoherence~\cite{kn:vion}. Understanding 
decoherence in a generic operating point may help in designing more flexible
gates, or to implement different strategies as computation with geometric phases~\cite{kn:nature}.

\begin{figure}[b]
\label{fig:threshold}
\includegraphics[width=0.5\textwidth]{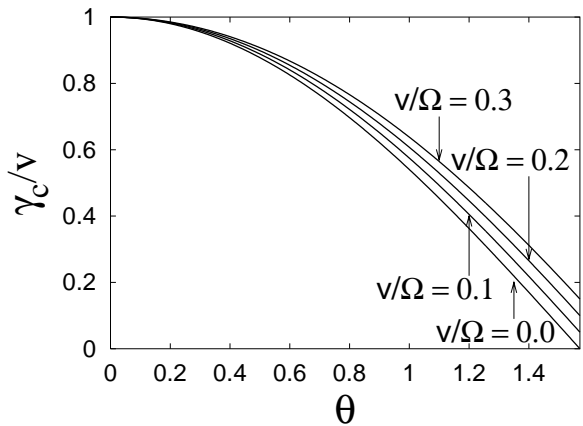}
\caption{
The threshold value for a BC behaving as weakly coupled depends on
the operating point}
\end{figure}
\subsection{Single BC}
First we consider Eq. (\ref{eq:adiabatico-prima-analisi-1}) and we notice that 
by changing the working point 
$\theta$, a strongly coupled BC may become weakly coupled and vice-versa.
For simplicity we let
$K T \gg \varepsilon_c$, then 
$\alpha \approx \sqrt{c^4 - g^2}$. The BC is weakly coupled if 
$\gamma \gg \gamma_c = (\Omega^\prime - \Omega)/c^2$.
Thus, the threshold value
decreases if we go from $\theta=0$ to $\theta=\pi/2$ (see Fig. ~\ref{fig:threshold}).

To get some insight in the problem of decoherence away from 
$\theta = \pi/2$ we consider $\Gamma(t)$ in the adiabatic regime 
Eqs. (\ref{eq:decay-coherences},\ref{eq:adiabatico-prima-analisi-1}).
\begin{figure}
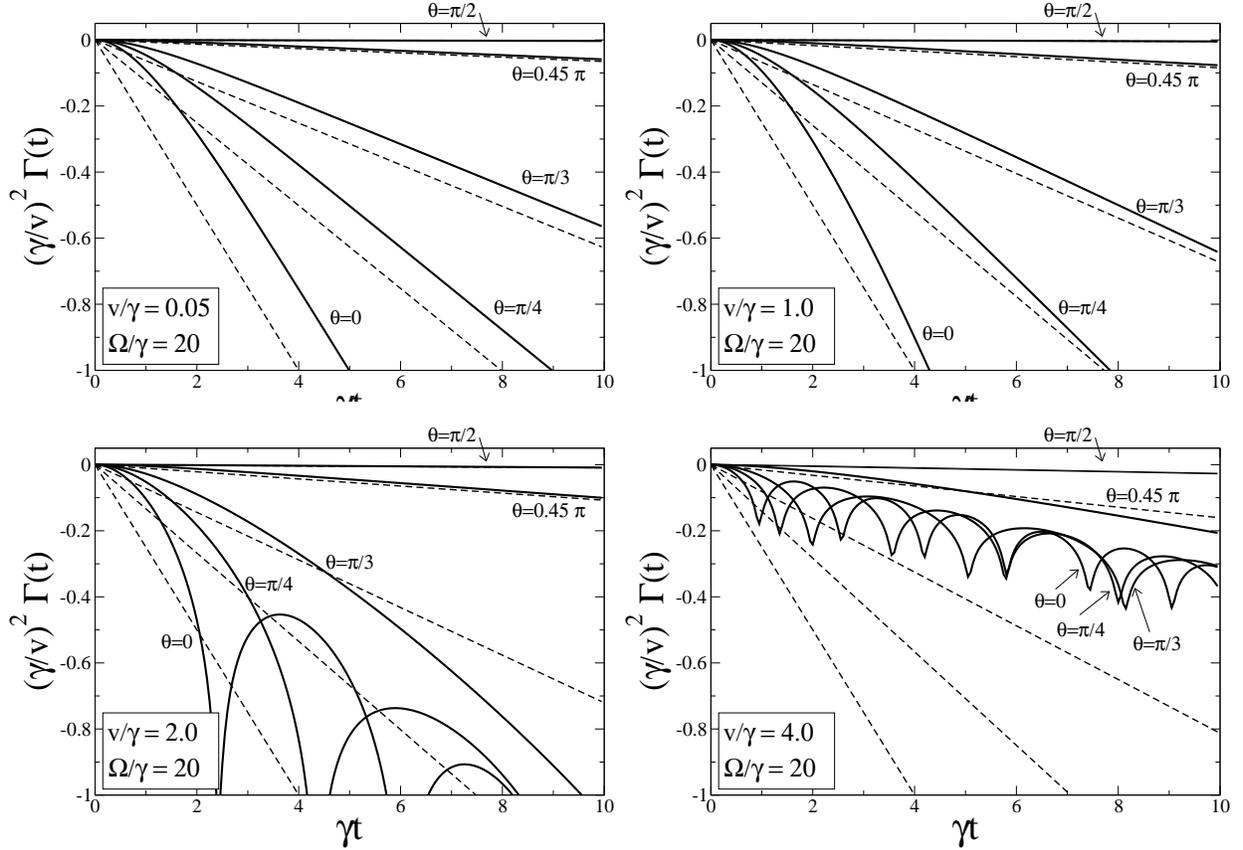

\includegraphics[width=0.49\textwidth]{palafig6.eps}
\includegraphics[width=0.49\textwidth]{palafig7.eps}\\
\includegraphics[width=0.49\textwidth]{palafig8.eps}
\includegraphics[width=0.49\textwidth]{palafig9.eps}
\caption{Reduced $\Gamma(t)$ (solid lines) plotted for 
various $v/\gamma \,=\, 0.05,\, 1.0,\, 2.0,\, 4.0$. In 
each panel curves correspond, from top to bottom, to 
$\theta = \pi/2, 0.45 \,\pi, \pi/3, \pi/4, 0$. 
For comparison $\Gamma_0(t)$ is also plotted 
(dashed lines). In the units used,  $\Gamma_0(t)$ is the same in each panel}
\label{fig:mmh}
\end{figure}
Results plotted in Fig.  \ref{fig:mmh} show $\Gamma(t)$ parametrized by 
$\theta$ for 
four values of the parameter $v/\gamma$. For reference we plot also 
(dashed lines) $\Gamma_0(t)$, Eq. (\ref{eq:decoherence-golden}).
The top left panel shows a weakly coupled BC for which $\Gamma(t)$ roughly follows 
$\Gamma_0(t)$. The other panels show 
BCs with $v \ge \gamma$ which turn from strongly coupled to weakly coupled 
 increasing $\theta \in [0,\pi/2]$. For these latter BCs
saturation is less effective in suppressing decoherence when the 
operation point is close to $\theta=\pi/2$, which may be an indication of the 
fact that their effect is more sensitive to deviations from the optimal point.

\subsection{$1/f$ Noise at the Optimal Point}
For a set of BCs 
the road-map for the reduced dynamics outlined in 
Sec. \ref{sec:reduced-dynamics} is not easily 
implemented by generalizing the master equation 
Eq. (\ref{eq:master-equation-sch}).
In \cite{kn:PRL} we used the Heisenberg equations of motion, where we 
factorize correlations between the qubit and the bands, and between different BCs. 
We are left with a set of $3(N+1)$ coupled differential equations, where $N$ is the 
total number of BCs. These approximations are expected to work for small $v_i$ 
but in effect they give accurate results for general values of $g_i$ even if 
$v_i / E_J$ is not very small, as we checked by comparing with numerical 
evaluation of the master equation for the qubit with one and two BCs.
We study decoherence at the optimal point $\theta=\pi/2$ ($\Omega = E_J$) 
via the Fourier transform of  $\langle \sigma_z (t)\rangle$.
We consider a set of weakly coupled BCs in the range 
$[\, 10^{-2}, 10 \,]\,  E_J$ 
which determines $1/f$ noise around the operating 
frequency with amplitude of the typical 
measured spectra~\cite{kn:zorin,nakamura-echo} 
extrapolated at GHz frequencies.
\begin{figure}
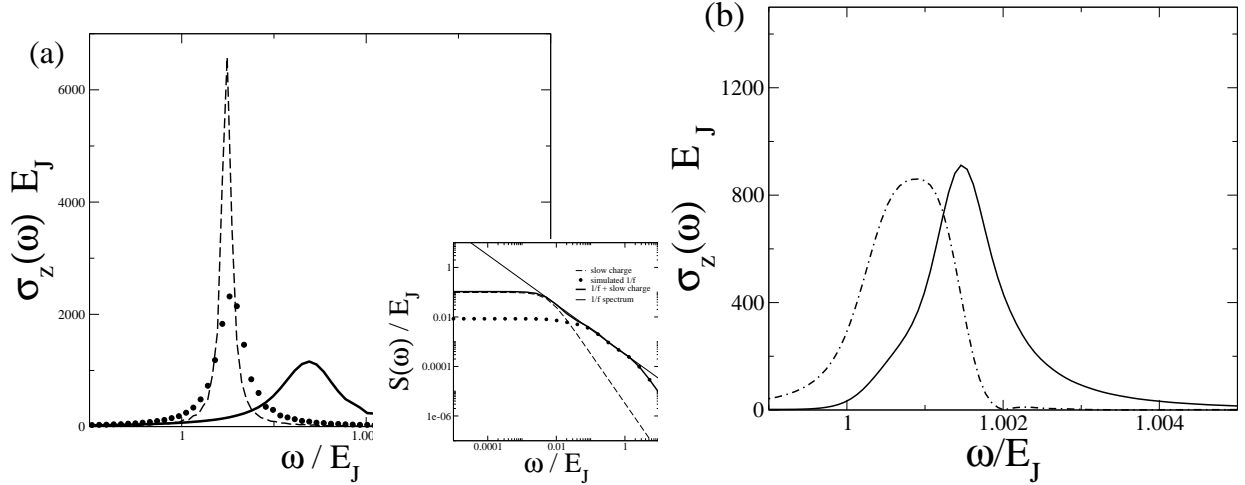

\includegraphics[width=0.45\textwidth]{palafig10.eps}
\hspace{-25mm}
{\includegraphics[width=0.22\textwidth]{palafig11.eps}}
\includegraphics[width=0.46\textwidth]{palafig12.eps}
\caption{(a) The Fourier transform $\sigma_z(\omega)$ 
for a set of weakly coupled BCs 
plus a single strongly coupled BC (solid line). The separate effect of 
the coupled slow BC alone ($g_0= 8.3$, dashed line) and of the 
set of weakly coupled BCs 
(dotted line), is shown for comparison.
In the inset the corresponding power spectra: notice that at $\omega = E_J$
the power spectrum of the extra charge alone (dashed line) is very small. 
In all cases the noise level at $E_J$ is fixed to the value 
${\mathrm S}(E_J) / E_J \approx 3.18 \times 10^{-4}$.
(b)
The Fourier transform $\sigma_z(\omega)$ 
for a set of weakly coupled BCs plus a strongly coupled BC  
($v_0 / \gamma_0 =61.25$) prepared in the ground (dotted line) or in the 
excited state (thick line) 
}
\label{figure:equil}
\end{figure}
Dephasing due to this set of BCs agrees with the weak coupling result 
Eq. (\ref{eq:decoherence-golden}).
Then we add a slower (and strongly coupled $g_0=v_0/\gamma_0=8.3$) BC, 
which should produce no effect according to Eq. (\ref{eq:decoherence-golden}), 
Fig.  \ref{figure:equil}a. We find that the strongly coupled BC alone 
determines a dephasing rate comparable with that of the weakly coupled
BCs and the overall dephasing rate is more than doubled.
This result shows that slower charges $\gamma \ll \Omega$ play a role in dephasing. 
Moreover  information beyond
$S(\omega)$ is needed, as we checked by showing that sets of charges with different $N$ and 
$v_i$ but the same $S(\omega)$ yield substantially different values of 
the decoherence rate. Decoherence is larger 
if BCs with $g \raisebox{-3pt}{\small $\,\gtrsim$} 1$ are present in the 
set. In summary these results show that Eq. (\ref{eq:decoherence-golden})
{\em underestimates} the effect of strongly coupled BCs and also that 
decoherence
at the optimal point may be substantial even if the $1/f$ spectrum does not 
extend up to frequencies $\sim E_J$.

If we further slow down the added BC we find that $\Gamma_{\phi}$
increases toward values $\sim \gamma_0$, the switching rate of the BC. 
This indicates that the effect of 
strongly coupled BCs on decoherence tends to saturate, in analogy with 
the results for pure dephasing.
In this regime, we observe also 
memory effects related to the initial preparation of the strongly coupled BC 
(see Fig.  \ref{figure:equil}b). 
Again we expect a dependence of dephasing on the protocol of the quantum gate.

\section{Conclusions}
In conclusion we have studied dephasing due to charge fluctuations in 
solid state qubits. For a fluctuator environment with $1/f$ spectrum memory 
effects and higher order moments are important so additional information
on the environment is needed to estimate dephasing. 
In this case the additional 
information of the environment needed depends on the protocol but often
reduces to a single parameter. A new energy scale emerges, the average coupling
$\overline{|v|}$ of the qubit with the BCs, which is the additional 
information needed to discuss single shot experiments (alternatively one 
should know the order of magnitude of $n_d$). 
For repeated experiments the relevant scale is instead 
$ \min \{\overline{|v|}, 1/t_m\}$ where $t_m$ is the overall
measurement time. We point out that these scales 
emerge directly from the study of the dynamics of the model, 
and not from further assumptions. 
Finally echo measurements are sensitive to the high-frequency
cutoff $\gamma_M$ of the $1/f$ spectrum. 

It is interesting to notice that some
result relative to an environment of quantum harmonic oscillators for
arbitrary qubit-environment
coupling may be obtained as a limit of the discrete environment in the 
semi-classical regime, and depend on the classical statistical properties
of an equivalent random process
with no reference to the quantum 
nature of the environment.

Our results are directly applicable to other implementations of solid state 
qubits. Josephson flux qubits~\cite{kn:tian00,kn:mooji} 
suffer from similar $1/f$ noise, originated from trapped vortices. Our model
applies if $\sigma_z$ represents the flux.
Also the the parametric effect of $1/f$ noise on the coupling energy of a 
Josephson junction~\cite{kn:clark} can be analyzed within our model, as long as
individual fluctuators do not determine large variations of $E_J$. In this
case it may be possible that the same sources generate both charge noise 
and  fluctuations of $E_J$. This can be accounted for by 
choosing the ``noise axis'' as the $\hat{z}$ axis. 

An important issue is to understand dephasing 
near the optimal operating points of the qubit~\cite{kn:vion}. 
Low-frequency noise can also be minimized by echo techniques, but the 
flexibility in the implementation of gates is greatly reduced. A possibility
is to implement quantum computation using Berry phases~\cite{kn:nature}, 
where the design of gates includes an echo procedure. Frequency shifts
can be calculated within our model but a reliable
analysis of the effect of $1/f$ noise is still missing. 

Finally we mention that the sensitivity of coherent devices may be used 
to investigate high frequency noise~\cite{kn:kouvenowen}. In particular 
an accurate matching between measured inhomogeneous broadening,  
echo signal and relaxation may give reliable information on the actual 
existence of BCs at $GHz$ and on the high-frequency cutoff of the $1/f$ 
spectrum.

\section*{Acknowledgments}
\noindent 
We thank M. Palma, R. Fazio, A. Shnirman, 
G. Sch\"on and C. Urbina for discussions 
which greatly sharpened the point of view presented in this work. 
Very useful discussion with J. Clarke, A. D'Arrigo, D. Esteve, F. Hekking, 
P. Lafarge, G. Giaquinta, M. Grifoni, D. van Harlingen, A. Mastellone, 
J.E. Mooji, Y. Nakamura, Y. Nazarov, F. Plastina, U. Weiss, 
D. Vion and A. Zorin are acknowledged.

\end{document}